# A Photonic Atom Probe Analysis of the Effect of Extended and Point Defects on the Luminescence of InGaN/GaN Quantum Dots

I. Dimkou, [1,*] J. Houard, [2] N. Rochat, [1] P. Dalapati, [2,†] E. Di Russo, [1,2,⊥] D. Cooper, [1] A. Grenier, [1] E. Monroy, [3] and L. Rigutti [2]

[1] Univ. Grenoble-Alpes, CEA, LETI, F-38000 Grenoble, France
[2] UNIROUEN, CNRS, Groupe de Physique des Matériaux, Normandie Université, 76000 Rouen, France
[3] Univ. Grenoble-Alpes, CEA, Grenoble INP, IRIG, PHELIQS, 17 av. des Martyrs, 38000 Grenoble, France

* ioanna.dimkou@cea.fr

† Present address: Research Center for Nano-Devices and Advanced Materials, Nagoya Institute of Technology, Nagoya 466-8555, Japan
⊥ Present address: Dipartimento di Fisica e Astronomia, Università di Padova, Via Marzolo 8, 35131 Padova, Italy

OrcID:
Ioanna Dimkou: 0000-0002-6865-3820
Névine Rochat: 0000-0003-3574-4424
Jonathan Houard: 0000-0001-7244-205X
Pradip Dalapati: 0000-0001-9021-726X
Enrico Di Russo: 0000-0003-3829-6567
David Cooper: 0000-0003-3479-4374
Eva Monroy: 0000-0001-5481-3267
Lorenzo Rigutti: 0000-0001-9141-7706

We report a correlative microscopy study of a sample containing three stacks of InGaN/GaN quantum dots (QDs) grown at different substrate temperature, each stack consisting of 3 layers of QDs. Decreasing the substrate temperature along the growth axis leads to the proliferation of structural defects. However, the luminescence intensity increases towards the surface, in spite of the higher density of threading dislocations, revealing that the QD layers closer to the substrate behave as traps for non-radiative point defects. During atom probe tomography experiments combined with in-situ micro-photoluminescence, it was possible to isolate the optical emission of a single QD located in the topmost QD stack, closer to the sample surface. The single QD emission line displayed a spectral shift during the experiment confirming the relaxation of elastic strain due to material evaporation during atom probe tomography.

# I. INTRODUCTION

III-Nitride LEDs have reached their maturity and are now a standard commercial product that finds wide application, for example in lighting and signaling. However, some unsolved problems remain. The efficiency droop that appears associated with Auger phenomena sets a performance limit [1,2], which is currently circumvented through electronics and increasing the number of LED chips needed to maintain the high efficiency. The indium distribution in the active quantum wells (QWs) plays a relevant role on the Auger recombination rate. The use of self-assembled quantum dots (QDs) as active media has been proposed as an approach to attenuate the droop [3,4]. On the other hand, the wall-plug efficiency is maximum for devices that emit in the blue range, but it decreases markedly for green LEDs, with a higher In content in the active region, therefore higher alloy inhomogeneities and higher density of misfit-related defects. These problems become even more acute in III-nitride based red LEDs, still in development [5]. It is hence important to perform correlated optical/structural studies at the nanoscale, to understand the role of defects on the performance of light emitters. The correlation of transmission electron microscopy (TEM) and laser-assisted atom probe tomography (APT) provides visualization of chemical inhomogeneities at the nanometer scale, and sheds some light on their correlation with the device optical performance. This method has been first used for InGaN/GaN QWs by Bennet et al. [6]. The first systematic study by Riley et al. [7] compared the correlated scanning transmission microscopy (STEM) and APT data obtained on different sets of InGaN/GaN QWs in nanowire LEDs with the spatially resolved cathodoluminescence (CL) spectra collected on other nanowires from the same sample. We have recently demonstrate the application of this technique to the study of the indium distribution in Stranski-Krastanov InGaN/GaN QD layers [8], attaining excellent correlation between the structural/chemical study, optical measurements and three-dimensional (3D) theoretical calculations of the electronic structure of the QDs.

In this work, we analyze the optical performance of 3 stacks of InGaN/GaN QDs grown at different substrate temperature, each stack consisting of 3 layers of InGaN QDs. Lowering the substrate temperature leads to an increased density of structural defects [9]. However, the luminescence efficiency of the sample does not correlate with the structural quality. On the contrary, higher emission intensity is observed in more defective low-temperature layers. This result reveals the predominant role of point defects in the luminescence properties of InGaN nanostructures, and highlights the relevance of inserting InGaN underlayers to bury non-radiative point defects. Thus, in the topmost and more defective QD layers, the signature of single QD emission is observed in APT specimens.

## II. EXPERIMENT

### A. Sample growth

The test structure was synthesized by plasma-assisted molecular beam epitaxy (PAMBE), and the growth process was monitored by reflection high-energy electron diffraction (RHEED). In order to explore the role of the growth temperature and material quality on the optical properties of InGaN QDs, we designed sample containing 3 stacks of 3 layers of self-assembled $In_xGa_{1-x}N$/GaN (2 nm/12 nm) QDs, separated by 70 nm of GaN [see schematic description in Fig. 1(a)]. This structure was deposited on a GaN-on-sapphire template and capped with 500 nm of GaN. At the beginning of each QD stack, we changed the growth temperature and adjusted the indium flow to compensate for the change in the indium desorption rate. The substrate temperature was calibrated by measuring the In desorption time, as described elsewhere [10]. It decreased monotonically during the growth, fixed at 615°C, 580°C and 550°C for the first, second and third QD stack, respectively. The active nitrogen flux was maintained constant, with conditions that provide a GaN growth rate of 0.52 monolayers per second (ML/s) under metal-rich conditions. For the generation of InGaN QDs, the Ga flux was fixed at 30% of the stoichiometric value, and the In flux was tuned close to the stoichiometry [11]. The InGaN growth time was 34 s, preceded by 5 s of In deposition to assure the In-rich growth. After deposition of the QD material, the growth was interrupted for 15 s, resulting in a rough surface morphology observed by RHEED. Then, the GaN barriers were deposited with a Ga flux slightly higher than the stoichiometric value. Before the growth of the following QD layer, the Ga excess accumulated at the GaN growth was consumed with active nitrogen during 20 s.

**B. Characterization methods**

STEM studies were performed in an aberration-corrected FEI Titan 80-300 microscope operated at 200 kV, using high-angle annular dark field (HAADF) and bright field (BF) detectors.

CL mapping was carried out in an Attolight CL microscope. The acceleration voltage was 10 kV and the beam current was ≈ 5 nA. The luminescence was collected through an integrated microscope objective (NA:0.7). By scanning the sample, the optical spectra of each pixel are recorded on a CCD camera through a dispersive spectrometer with 400 mm focal length (grating for spectral measurements: 150 grooves/mm blazed at 500 nm with ×2 binning; grating for CL mapping: 600 grooves/mm blazed at 300 nm with ×8 binning).

Secondary ion mass spectrometry (SIMS) investigations were performed in a SIMS magnetic selector instrument from CAMECA [12], using $^{133}$Cs$^+$ as primary beam, accelerated at 2.0 kV, and analyzing a 175×175µm spot.

Needle-shaped tips required for APT and lamella specimens for TEM were prepared by scanning electron microscope/focused ion beam (SEM/FIB) using a Zeiss Nvision40 and a Zeiss Crossbeam550, respectively. To extract the region of interest, the as-grown sample was previously coated with a 500-nm-thick Pt layer using a GATAN PECS system. Lift-out procedures were carried out with Ga+ ions accelerated 30 kV. Annular milling and lamella thinning procedures were performed with an acceleration voltage of 30 kV in a first stage, and then 2 kV for the final milling and cleaning stages, to minimize the thickness of the amorphized region [13,14].

The samples were analyzed in a homemade laser-assisted tomographic atom probe system coupled with a micro-photoluminescence (µPL) bench, also referred to as a photonic atom probe (PAP) [15,16]. The evaporation of the specimen is triggered by pulsed laser (pulse width $\approx$ 150 fs, repetition frequency = 400 kHz, average power = 14 µW i.e. 35 pJ/pulse) emitting at 260 nm and focused on a spot with a diameter around 1.5 µm. The atom probe system was equipped with an 8-cm-diameter multichannel plate/delay line detector (MCP/DLD). The flight length is $\approx$ 22 cm, corresponding to a field of view of +/- 10° on the detector and of +/-15° on the tip. The detection rate was constant $\varphi \approx 0.0001$ event/ pulse. The µPL signal was collected through a grating spectrometer with 320 mm focal length equipped with a liquid nitrogen-cooled CCD camera. The spectral resolution was approximately 0.3 nm.

## IV. RESULTS AND DISCUSSION

Figure 1(b) presents a HAADF-STEM image of the heterostructure where the three InGaN/GaN QDs stacks can be identified: InGaN is located in discontinuous horizontal lines with bright contrast, whereas GaN has darker contrast due to the difference in atomic number. The contrast fluctuations in the InGaN layers confirm the presence of InGaN quantum dots. Threading dislocations (white vertical lines) propagating from the underlying GaN substrate are outlined with yellow arrows. In the proximity of these initial defects, additional threading dislocations appear at the GaN/InGaN interface. This points to the generation of interfacial misfit dislocations, which thread upward [17], a well-known phenomenon in In-rich InGaN/GaN heterostructures. The local strain in In-rich areas may cause the decomposition of (a+c)-type threading dislocations to produce an a-type and a c-type dislocation threading towards to the surface. The a-type component of the dislocation bends to an interfacial direction giving rise to a misfit dislocation with Burgers vector b=$\frac{1}{3}$<11-20> in the {0001} slip plane. When the misfit dislocation attains an In-poor area, the local change in strain provides the driving force for the dislocation to climb band to the [0001] direction. The decrease of temperature during the growth is an important factor promoting the generation of misfit dislocations and the general decline of material quality [18–20], in agreement with our STEM observations. Moreover, it is known that In can incorporate along the defects, forming In nanoparticle arrays extending along the growth direction [21].

To confirm the presence of In in the ultrathin InGaN QD layers, and assess the defect-promoted diffusion of In along the growth axis, we have used secondary ion mass spectroscopy (SIMS). Results are reported in Fig. 1(c). All the QD layers are well identified in the profile, and the three bottommost and the three topmost layers seem to have roughly the same amount of In. On the contrary, the 3 layers in the middle present a slightly higher total amount of In. The measurement of the In concentration in the QD layers is not quantitative here due to the small size of the nanostructures and their three-dimensional nature. In the GaN material within the QD layers, the In concentration goes down below the detection limit of the system ($\approx 4 \times 10^{17}$ cm$^{-3}$).

Figure 2(a) presents a low-temperature (T = 5 K) CL map (wavelength window = 395-420 nm) of the STEM lamella specimen (thickness of 100-120 nm), compared with a section of the HAADF-STEM image, in Fig. 2(b), to visualize the location of the QD stacks. In Fig. 2(a), most of the CL intensity originates in the two topmost QD stacks, being significantly higher in the topmost stack, grown at the lowest substrate temperature. This trend was confirmed at several points of the specimen. Figure 2(c) presents the CL spectra extracted from slices of Fig. 2(a), moving from the GaN substrate (1), crossing the 3 InGaN/GaN QD stacks (2-6), to the middle of the GaN cap layer (7,8). All the spectra show a narrow emission line at 357 nm, characteristic of GaN. The spectra recorded along the QD region display also a broad emission that peaks at 387-402 nm, which is assigned to recombination in the QDs. Along the growth axis, the peak emission wavelength shifts only by 15 nm, which is smaller than the emission linewidth (spectra 3, 4 and 5 show QD emission linewidths of 29 nm, 33, nm and 17 nm, respectively). If we keep in mind that the CL measurements were performed on a lamella specimen, which means that the electron beam can only excite a few dots, the spectral shift and variation of the emission are explained by the expected dot-to-dot variations of morphology and indium composition.

In order to analyze the indium distribution in the InGaN/GaN QDs and the luminescence at the scale of a single emitter, field-emission tip specimens were fabricated and characterized by La-APT. FIB prepared tips were first observed in TEM, as illustrated in Fig. 3(a). The tips were then introduced in the PAP system. To determine the best experimental conditions to perform simultaneous µPL and APT characterization, we first measured *in-situ* µPL as a function of the laser pumping power at 28 K, as shown in Fig. 3(b). During these preliminary measurements, we used a repetition rate of 3.8 MHz, adjusting the laser power in the range of 25 to 600 µW. In these spectra, it is possible to identify the emission of GaN at 356 nm and a narrow line around 390 nm, which is assigned to InGaN QDs. As previously observed [8], the luminescence from GaN increases monotonously with the pumping power, but the QD emission intensity saturates [see Fig. 3(c)]. More importantly, the full width at half maximum (FWHM) of the main QD line shows a drastic increase for excitation power higher than 200 µW [see Fig. 3(d)], which could point to an increase of the temperature. Considering this observation, we decided to perform the La-APT experiment with a laser pulse energy corresponding to 140 µW, to prevent thermal effects or artefacts associated with many-body phenomena.

During the APT measurements coupled with µPL, the laser frequency was reduced to 340 kHz for the reasons that are well explained in ref. [22]. The tip was kept at 40 K. In the experiment described in Fig. 4, the tip contained originally the three InGaN/GaN QDs stacks. Fig. 4(a) presents PL spectra recorded at various moments of the evaporation process, and Fig. 4(b) shows 3D APT reconstructions of the indium signal where we can clearly identify the QD layers. The three images in (b) illustrate what was left on the specimen when the spectra #4, #5 and #6 in (a) were recorded. The spectra did not display major variations during the evaporation of the two topmost QD layers of the first QD stack (spectra #1 → #5). As explained before, the peak at 356 nm is related to the carrier recombination in GaN, and the narrow line at 398 nm is assigned to the carrier recombination in the InGaN QDs. The FWHM, around 1.5 nm confirms that this peak comes from a single dot, comparing to the literature [23–25]. During the evaporation of the third QD layer (spectrum #6), the QD emission disappeared. This result confirms that the emitting single dot was located in the third top most QD layer. Note that the QD signal disappeared even if there were still InGaN layers, which means that either the other InGaN layers below do not contain any dot or non-radiative recombination processes dominate.

The evolution of the peak wavelength and FWHM of the QD line during the evaporation of the tip is described in the Fig. 4(c). The QD emission line presents a red shift of around 2 nm during the APT process of the first 3 topmost QD layers. This shift is mostly explained by the relaxation of the misfit strain during the evaporation, since the possible contribution of the stress induced by the external electric field when the dot approaches the surface of the tip [25] is very small in this system [8]. Note that the difference of the voltage applied to the InGaN/GaN tip from the beginning to the end of the experiment (spectra #1 → #5) is in the range of $\Delta V_{DC}$ = 1 kV ($V_{DC}$ = 4.5-5.5 kV), which is not enough to justify such spectral shift. The FWHM of the emission, indicated as an error bar, does not present a clear trend.

Figure 5(a) shows the indium site fraction map obtained through the analysis of the whole tip. The InGaN layers contain In-rich regions with a typical in-plane diameter of around 15 nm and typical In-site fraction ≈ 10%, on a thin, discontinuous wetting layer, with an In-site fraction ≈ 2.5%. The shape and size of these In-rich regions correspond to InGaN QDs rather than to random alloy fluctuations [30, 35].

The distribution of electric field in the tip can cause artifacts in the measurement of alloy composition [27]. To assess the relevance of these effects, Fig. 5(b) illustrates the spatial distribution of the $Ga^{2+}/Ga^{+}$ charge state ratio. This map allows the correlation between surface electric field and measured composition visualized at a microscopic scale. The effective electric field can be estimated by the ratios of different charge states of given species through Kingham's post-ionization model [28]. The intense electric field is responsible for the evaporation of single charged ions from the tip surface. However, post-ionization [28] can occur close to the tip surface. When an ion reaches a critical distance from the tip surface (several angstroms), additional electrons can be transferred from the ion to the tip surface by quantum tunneling effects [29]. Comparing the Fig. 5(a) and (b), we can suggest that there is a certain correlation between the composition and the surface field. The high field region (up to 0.015) is associated with an In III-site fraction about 10 %. Comparing our values with those from literature [27,30–33], the highest field recorded here is sufficiently low to neglect compositional biases and to ensure accuracy of the measured composition.

Plane-views of the indium composition of the second topmost (outlined with a blue rectangle in Fig. 5(a)) and the first bottommost InGaN QD layer (outlined with an orange rectangle in Fig. 5(a)) are presented in Figs. 5(c-d) and (e-f), respectively. Note that the scales of Figs. 5(c) and (e) are the same, pointing to relatively similar indium distribution and QD diameter, in the range of 10-20 nm. Moreover, the spatial distribution of the $Ga^{2+}/Ga^{+}$ charge state ratio in Figs. 5(d) and (f) are also similar. From this chemical analysis, the 3 stacks of QD layers are quite similar in terms of size and composition, which does not justify major changes in the radiative properties.

Note that Figs. 5(a), (c) and (e) are slices of the tip reconstruction with a thickness of 10 nm in the direction perpendicular to the page. The indium concentration is calculated by integrating the In and Ga atoms in the depth of the slice. As the base diameter of the QDs is in the range of 10-20 nm and the height is in the range of 2 nm, the In content in the dots in the plane-views (Figs. 5(c) and (e)) is strongly underestimated due to the surrounding GaN. This is the reason why the In-site fraction for the in-plane is only ≈ 5%, compared to 10% in Fig. 5(a).

Coming back to the structural properties of the sample, there is an important degradation of the structural quality along the growth axis due to the progressive reduction of substrate temperature during the growth. These results, illustrated in Fig.

1(b), seem to be in contradiction with the systematic enhancement of the luminescence intensity along the growth axis. It is known that the potential fluctuations generated by inhomogeneities in the In content can reduce non-radiative recombination in InGaN layers [34,35], to the point of suppressing the non-radiative effects associated to dislocations [36]. However, the effect of carrier localization in In-rich areas should be visible from the first QD layer. Therefore, enhancement of the luminescence along the growth axis points to the fact that non-radiative processes in the layers close to the substrate are dominated by point defects. This is consistent with recent reports of points defects as a main source of degradation of the luminescence of InGaN/GaN quantum well structures [37–39]. Several groups have demonstrated that such defects can be buried by depositing InGaN layers before the growth of the active region [37,38,40]. In our case, the first QD layers in the structure play the role of InGaN underlayers trapping non-radiative point defects. The filtering efficiency of such layers is high enough to resolve single QD luminescence in the topmost stack, in spite of the high density of threading dislocations.

## VI. CONCLUSIONS

To summarize, we have performed a correlative microscopy study of 3 stacks of 3 layers of self-assembled $In_xGa_{1-x}N$/GaN (2 nm/12 nm) QDs, separated by 70 nm of GaN, and each of the three stacks grown at different substrate temperature. Although lowering the substrate temperature leads to an increased density of structural defects, we observed higher emission intensity in the topmost layers, grown at lower temperature. This result highlights the relevance of inserting InGaN underlayers to bury non-radiative point defects. During atom probe tomography measurement of the topmost QD layers, it was possible to resolve one single QD emission line. The single QD line displayed a spectral shift during the experiment confirming the relaxation of elastic strain due to material evaporation.

## ACKNOWLEDGMENTS

This work was co-funded in the framework of RIN IFROST, and CPER BRIDGE projects by European Union with European Regional Development Fund (ERDF) and by Region Normandie. Partial support from the French National Research Agency, within the project "INMOST" (ANR-19-CE08-0025) is acknowledged. This study was performed on the NanoCharacterisation PlatForm (PFNC), and supported by the "Recherches Technologiques de Base" Program of the French Ministry of Research.

---

**Figure Captions**

**FIG. 1.** (a) Schematic of the InGaN/GaN QD structure. (b) HAADF-STEM image of the sample observed along the [10-10] zone axis (lamella specimen thickness 100-120 nm). Yellow arrows indicate threading dislocations propagating from the substrate. (c) SIMS depth profile for indium.

**FIG. 2** (b) Low temperature (5 K) CL mapping (spectral window = 395-420 nm) of the STEM lamella imaged in Fig. 1, containing the 3 stacks of InGaN/GaN QDs, compared with (b) a section of the HAADF-STEM image of the same specimen. (c) CL spectra extracted from different areas of (b), moving from the GaN substrate and through InGaN QDs towards the surface (#1 → #8).

**FIG. 3.** (a) TEM image of the tip after FIB preparation. (b) Micro-PL spectra before La-APT evaporation of a tip specimen containing InGaN/GaN QDs. Spectra recorded after excitation with different levels of laser power were normalized to the maximum of the QD emission and vertically shifted for clarity. The experiment was performed at 28 K and the pulse rate and wavelength of laser were 300 kHz and 260 nm, respectively. Variation of (c) the peak PL intensity and (d) FWHM of the lines assigned to the QDs and to GaN, as a function of the laser power.

**FIG. 4.** (a) *In-situ* μPL measurements performed during the La-APT evaporation of InGaN/GaN QDs. Evaporation takes place from 1 to 6. The experiment was performed at 40 K. (b) Quantitative indium-site 3D APT reconstruction of the tip. The images describe what was left of the specimen when recording the spectra 4, 5 and 6 in (a). (c) μPL peak wavelength of the QD signal in the spectra presented in (a). The FWHM is indicated as error bars.

**FIG. 5.** (a) Indium-site map calculated from the APT 3D reconstructed volume illustrated in Fig. 4(b). (b) Ga charge-state ratio ($Ga^{++}/Ga^{+}$) revealing the field distribution. The results are shown for a side view (perpendicular to the c axis). (c) Top view map (along the c axis) of the indium content and (d) Ga charge-state ratio ($Ga^{++}/Ga^{+}$) in the second topmost InGaN QD layer, outlined with a blue rectangle in (a). (e) Top view map (along the c axis) of the indium content and (f) Ga charge-state ratio ($Ga^{++}/Ga^{+}$), outlined with an orange rectangle in (a). Note that the color scales in (c) and (e) are the same, and those in (d) and (f) are also the same. The indium content shown in (a), (c) and (e) are relative to a 10 nm thick slice in the direction perpendicular to the page.

**FIG. 1**

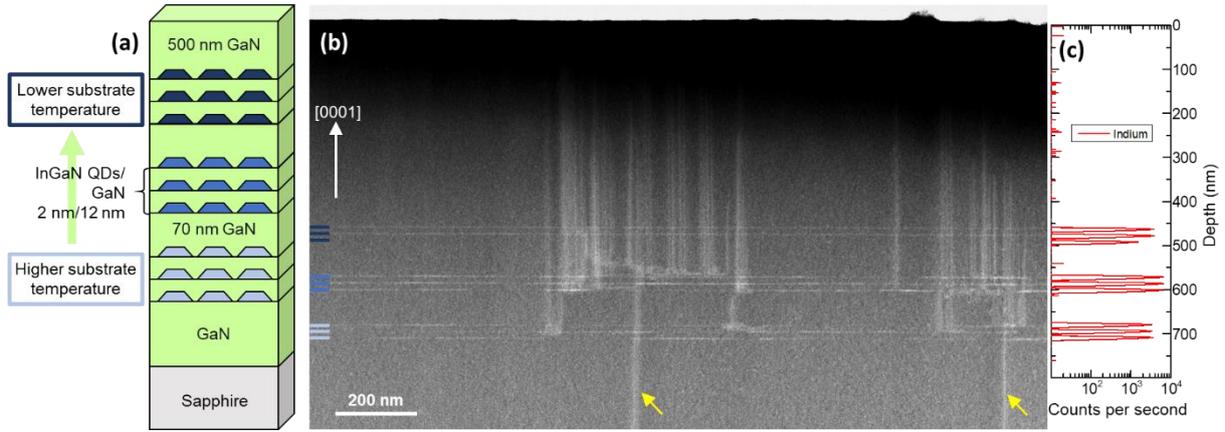

**FIG. 2**

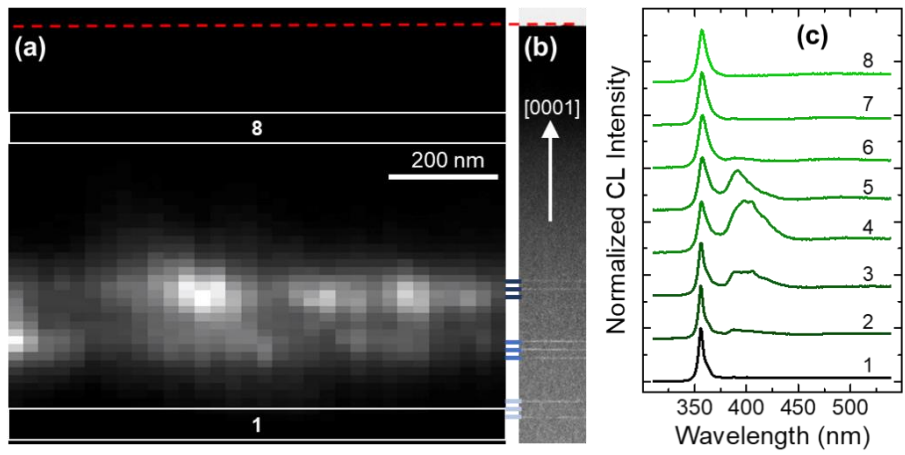

**FIG. 3**

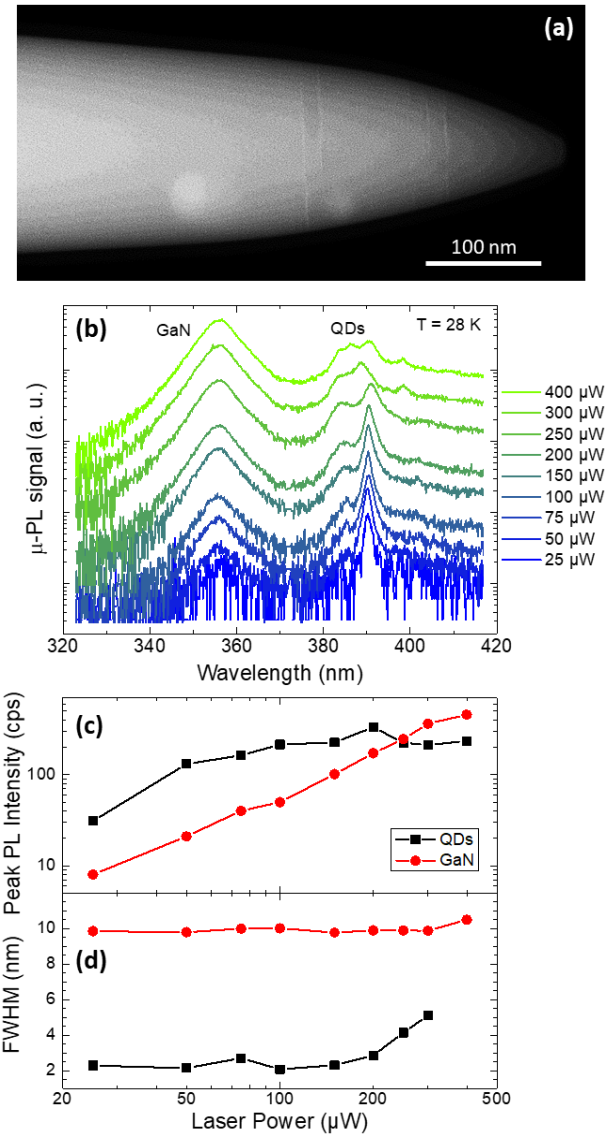

**FIG. 4**

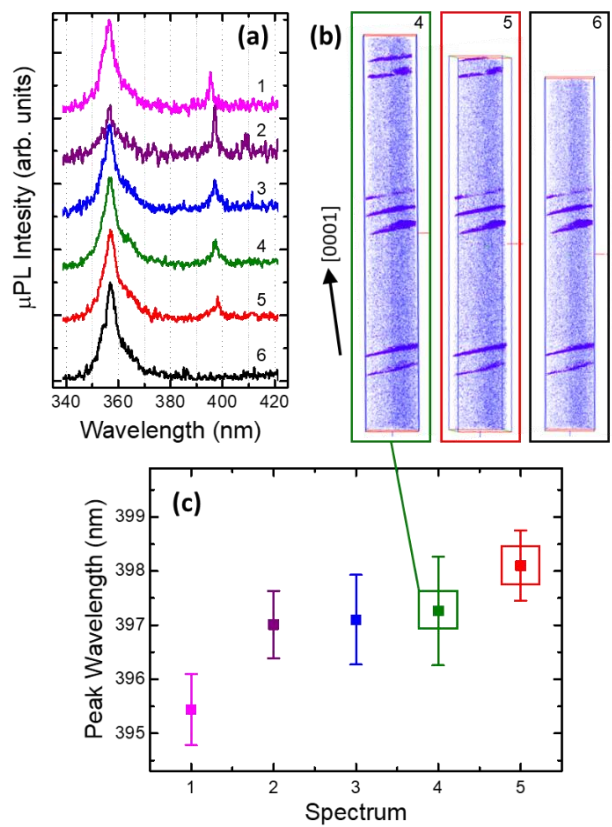

**FIG. 5**

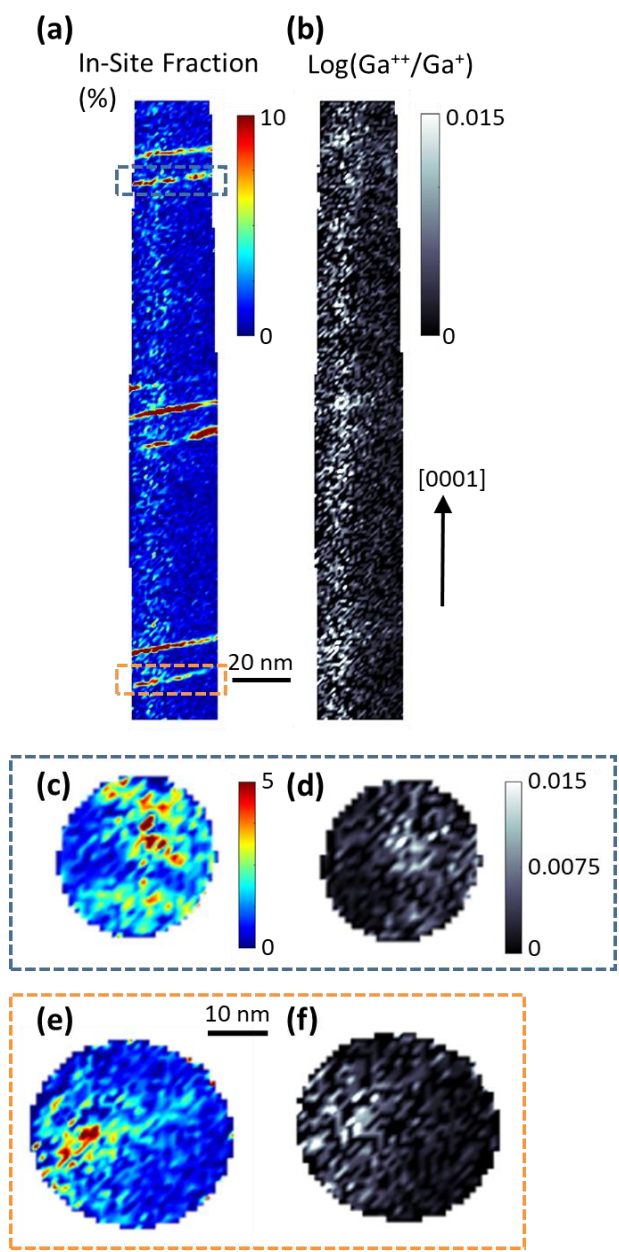